\documentclass{article}
\usepackage{graphicx}  
\usepackage{amsmath}   
\usepackage{amssymb}   
\usepackage{bm} 
\usepackage{dcolumn}
\usepackage{color}
\usepackage{mathrsfs}
\usepackage{amsfonts}
\usepackage{varioref}
\RequirePackage[colorlinks,citecolor=blue,urlcolor=magenta,linkcolor=blue]{hyperref}
\def\be{\begin{equation}}
\def\ee{\end{equation}}

\DeclareMathOperator{\cosech}{csch}
\labelformat{section}{Section #1} 
\labelformat{subsection}{Section #1} 
\labelformat{subsubsection}{Section #1}
\labelformat{subsubsubsection}{Section #1}
\labelformat{equation}{Eq.~(#1)} 
\labelformat{figure}{Fig.~#1} 
\labelformat{subfigure}{Fig.~\thefigure#1} 
\labelformat{table}{Tab.~#1} 
\labelformat{appendix}{Appendix #1}

\title{\bf Unruh-DeWitt detector responses for complex scalar fields   in de Sitter spacetime}
\author{Md. Sabir Ali$^1$\footnote{sabir.ali@iitrpr.ac.in},\,\,\, Sourav Bhattacharya$^1$\footnote{sbhatta@iitrpr.ac.in}~~and~~Kinjalk Lochan$^2$\footnote{kinjalk@iisermohali.ac.in}\\
$^{1}$\small{Department of Physics, Indian Institute of Technology Ropar, Rupnagar, Punjab 140 001, India}\\
$^{2}$\small{Department of Physical Sciences, IISER Mohali, Manauli 140 306, India} }
\begin{document}
\maketitle
\begin{abstract}
\noindent
We derive the response function for a comoving, pointlike  Unruh-DeWitt particle detector coupled to a complex scalar field $\phi$, in the $(3+1)$-dimensional cosmological de Sitter spacetime. The field-detector coupling is taken to be proportional to $\phi^{\dagger} \phi$. We address both conformally invariant and massless minimally coupled scalar field theories, respectively in the conformal and the Bunch-Davies vacuum.   The response function integral for the massless minimal complex scalar, not surprisingly, shows divergences and accordingly we use suitable  regularisation scheme to find out well behaved  results. The regularised result also contains a de Sitter symmetry breaking logarithm, growing with the cosmological time. Possibility of extension of these results with the so called de Sitter $\alpha$-vacua is discussed. While we find no apparent problem in computing the response function for a real scalar in these vacua,  a complex scalar field is shown to contain some possible ambiguities in the 
detector response. The case of the minimal and nearly massless scalar field theory is also briefly discussed. 
\end{abstract}
\vskip .5cm


\newpage
\tableofcontents
\section{Introduction}\label{intro}

An Unruh-DeWitt detector is conventionally a  point particle (like an atom) that can couple to a quantum 
field. The detector has internal discrete energy levels which, along with the field  may be excited/de-excited to higher/lower  levels. Such (de-)excitation depends  upon the trajectory of the detector, the field-detector coupling and also the particular initial and final states we are looking into.  One particularly interesting quantity is the response function of the detector, representing the rate of  quantum  transitions occurring per unit proper time along detector's trajectory. The associated  quanta are not necessarily actual created particles which may give rise to flow of energy and momentum, but instead they may be an outcome of application  of the external energy required to maintain detector's particular trajectory. We refer our reader to~\cite{Birrell:1982ix} and references therein for a discussion. The response functions for an Unruh-DeWitt detector have been investigated in various contexts  in the flat spacetime. This includes various non-inertial trajectories, e.g.~\cite{Svaiter:1992xt, Zhu:2007nm, Doukas:2013noa, Sriramkumar:1994pb, Sriramkumar:1998ms, Kothawala:2009aj, Gutti:2010nv, Barbado:2012fy} (also references therein).  See also~\cite{Ostapchuk:2011ud, Hu:2012jr, Menezes:2015iva} and references therein for dynamics of entangled detectors interacting with quantum fields.

The de Sitter spacetime is physically very well motivated in the context of the early inflationary  as well as the current universe. It has gained considerable attention in the context of the Unruh-DeWitt detector model. The earliest of such discussions can be seen in~\cite{Birrell:1982ix} and references therein, where it was shown that the response function for a comoving detector in the cosmological de Sitter spacetime for a conformal scalar in a conformal vacuum is thermal, although there is no actual particle creation in this scenario. This analysis was later extended in various directions, including scalar fields without conformal symmetry, in the static de Sitter coordinate and also quite extensively in the context of quantum entanglement and decoherence, e.g.~\cite{Garbrecht:2004du, Garbrecht:2004ui, Fukuma:2013uxa, Singh:2013pxf, Tian:2013lna, Ahmadzadegan:2014pva, Tian:2014jha, Lynch:2015lra, Rabochaya:2015aza, Liu:2016ihf, Tian:2016uwp, Kukita:2017tpa, Huang:2017yjt, Stargen:2017xii, Liu:2018zod, 
Yamaguchi:2018cqw, Farias:2019lls, Kaplanek:2019dqu, Kaplanek:2019vzj, Hotta:2020pmq} and references therein.  \\

In this work we compute at the leading order of the perturbation theory, the response functions  of a comoving Unruh-DeWitt detector for complex scalar fields, for both conformally symmetric and massless minimal cases. It is well known that a massless minimal scalar can be a very good candidate for the inflaton.  We also wish to discuss the possibility of extending  these results in the context of the de Sitter $\alpha$ vacua~\cite{Allen:1985ux, Bousso:2001mw,  Einhorn:2002nu, Banks:2002nv, Collins:2003zv, Einhorn:2003xb, Collins:2003mj, Collins:2004wi, Collins:2004wj, Collins:2007jc, Collins:2008xk}. Since the field-detector coupling must be hermitian,  for a complex scalar it
must be at the simplest non-trivial form proportional to $\phi^{\dagger} \phi$,\footnote{One can also envisage a Hermitian linear coupling with either $ \phi_R = \phi+ \phi^{\dagger}$ or $\phi_I = i(\phi^{\dagger}-\phi)$, which technically will be  no different from the monopole coupling of a real scalar field.  } unlike the case of a real scalar~\cite{Birrell:1982ix}.  Apart from considering this as a theoretical model of handling Dirac fermionic fields, such quadratic couplings can also be expected emerging in low energy effective theories of some interacting theories where a scalar describes composite particles at low energies \cite{Hinton:1984ht, Hummer:2015xaa}. The same is true for any complex field like a Dirac fermion. Due to such coupling, one obtains product of two Wightman functions in the integral of the response function and thus perhaps not unexpectedly, one encounters divergences needing suitable regularization. We refer our reader for discussions on the Unruh-DeWitt detector models for complex 
fields in the Rindler space  including entanglement dynamics to~\cite{Hinton:1984ht,    Hummer:2015xaa, Takagi:1984cd,  Bessa:2012fs, Zhou:2012gu, Harikumar:2012ff, Louko:2016ptn, Sachs:2017exo, Sachs:2018trp}.  Expectedly, in curved spacetime, the issue of the non-linear interaction between the detector and the field as well as the corresponding divergences may become much more relevant receiving curvature contributions. In order to analyse the curvature effects, we, therefore, consider the response of the Unruh-DeWitt detector in a maximally symmetric spacetime with constant curvature. This study can be considered as  a theoretical approach of handling the divergences appearing in more realistic non-linear couplings in various kind of fields in de Sitter spacetime. We refer our reader to~\cite{Prokopec:2006ue} and references therein for discussions on loop effects with a massless minimal complex scalar in the context of scalar quantum electrodynamics in de Sitter spacetime.
 
A quantum field in the de Sitter spacetime may inherit many inequivalent vacua, depending upon its symmetry structure. The Bunch-Davies vacuum, for example,  is suited for describing early time inflationary modes \cite{Parker:2009uva} according to comoving observers. The one parameter family of de Sitter invariant $\alpha$-vacua~\cite{Allen:1985ux} on the other hand, may be interesting in the context of the trans-Planckian physics. For example, discussion on such vacua in the context of the so called trans-Planckian censorship conjecture can be seen in~\cite{Bedroya:2020rmd} and references therein. Since an $\alpha$-vacuum can be expressed as a squeezed state over the Bunch-Davies states,
one can expect nontrivial feature in the detector response function with respect to such a vacuum. Another curious
point to note is, unlike the Bunch-Davies case, the timelike parameter that defines  a positive frequency $\alpha$-mode will not be the cosmological time, which is the proper 
time along a comoving detector's trajectory. 
Also, it was reported earlier in certain contexts that such vacua may not be good candidates to do perturbation theory, owing to their inherent non-local characteristics~\cite{Einhorn:2002nu, Banks:2002nv, Collins:2003zv, Einhorn:2003xb, Collins:2003mj, Collins:2004wi, Collins:2004wj, Collins:2007jc, Collins:2008xk}.   While we wish to
reinforce this argument in this paper, one possibly cannot completely rule these
vacua  out based upon the limitations offered by the perturbative approaches, for it is possible that they should be treated via some (hitherto unknown)
non-perturbative techniques. Not surprisingly, the analysis of $\alpha$ vacua has invited enough  attention and there has been a significant effort devoted
towards isolating their signatures in inflationary paradigm \cite{Collins:2003zv,deBoer:2004nd,  Kanno:2014lma}.
 
 The paper is summarised as follows. In the next section we briefly review the basic set up for the Unruh-DeWitt detector model. Using this, we discuss in \ref{free} the  response functions for a {\it real} conformal and massless minimal scalar. In \ref{s3}, we discuss complex conformal and massless minimal scalars respectively in the conformal and the Bunch-Davies vacua. For the latter in particular, suitable regularization scheme is employed in order to find out well behaved result. This regularization involves in particular, adding a fictitious conformal scalar field with divergent detector-field interaction, in order to cancel a divergence appearing in the response function.  This result also possesses a de Sitter symmetry breaking logarithm growing with time analogous to the infrared secular growth, reported earlier in e.g.~\cite{Brunier:2004sb, Karakaya:2019vwg, Baumgart:2019clc, Woodard:2014jba,
 Moreau:2018lmz}. However, such term is absent for the case of a real scalar~\cite{Garbrecht:2004du}. In \ref{alpha-com}, we discuss generalisation of complex fields'   results to the $\alpha$-vacua and we point out some possible ambiguities. The case of the nearly massless minimal scalar is briefly mentioned in \ref{nmm}. Finally we conclude in \ref{disc}. Even though we stick to the first order perturbation theory throughout, we argue in \ref{disc} that the response function for real scalars in the $\alpha$-vacua can be defined and computed  at any arbitrary order of the perturbation theory.

\section{The setup} \label{setup}
Following~\cite{Birrell:1982ix, Allen:1985ux, Bousso:2001mw},  we  briefly review below the basic set up we use in this paper, for the sake of completeness. 

The de Sitter metric in the spatially flat cosmological coordinates in $(3+1)$-dimensions reads
\begin{eqnarray}
ds^2= - dt^2 + e^{2H t} \left(dx^2+dy^2+dz^2\right),
\label{d1}
\end{eqnarray}
where $H =\sqrt{\Lambda/3}$ is the Hubble constant. Defining the conformal time, $\eta = - e^{-Ht}/H$, the metric  takes a conformally flat form,
\begin{eqnarray}
ds^2= \frac{1}{H^2 \eta^2}\left[- d\eta^2 + dx^2+dy^2+dz^2\right].
\label{d2}
\end{eqnarray}
The generic free action for a  real scalar field  reads, 
$$S= -\frac12 \int \sqrt{-g}\, d^4 x\, \left[  (\nabla_{\mu}\phi)(\nabla^{\mu}\phi) + m^2 \phi^2+ \xi R \phi^2 \right],$$
whereas for a complex scalar field it reads,
$$S= -\int \sqrt{-g}\, d^4 x \,\left[ (\nabla_{\mu}\phi^{\dagger})(\nabla^{\mu}\phi) +m^2 |\phi|^2+ \xi R |\phi|^2 \right].$$
We are chiefly interested in two  cases here : a) a conformal scalar ($m^2+\xi R = R/6$) and b) a massless minimal scalar ($m^2+\xi R =0$). The case of a nearly massless and minimal scalar will be briefly discussed in \ref{nmm}. We shall set below $\hbar =1=c$. \\

We shall use the formalism of particle detectors in curved spacetime discussed in e.g.~\cite{Birrell:1982ix} and references therein. Let us first discuss a real scalar field theory.  The simplest coupling of this field with a pointlike detector (e.g. an atom) is taken as,
$${\cal L}_{\rm int}= g \mu (\tau) \phi (x(\tau)),$$
where $g$ is a coupling constant and $\mu$ is the monopole moment operator of the detector. In the Heisenberg picture, $\mu(\tau)= e^{iH_0 \tau }\,\mu\, e^{-iH_0 \tau}$, where $H_0$ is the free Hamiltonian of the detector, and  $\tau$ is the proper time along detector's trajectory. We shall specialise to a comoving trajectory and hence will take detector's spatial points to be fixed. 

Thus the first order matrix element for the field-detector combined system to make a transition from an initial state $| i\rangle$ to a final state $| f \rangle$ is given by
\begin{eqnarray}
{\cal M}_{fi} = ig \,\langle E | \mu | E_0 \rangle \int_{\tau_i}^{\tau_f} d\tau e^{- i (E-E_0) \tau } \langle \phi_f | \phi(x(\tau)) | \phi_i\rangle,
\label{d4}
\end{eqnarray}
where we have taken $| i\rangle = | E_0 \rangle \otimes|  \phi_i \rangle$ and $| f\rangle = | E \rangle \otimes | \phi_f \rangle$, and $E_0$ and $E$ are respectively the energy eigencvalues of the detector in these states. The transition probability is given by 
$$|{\cal M}_{fi}|^2=g^2 \,|\langle E | \mu | E_0 \rangle|^2 \int_{\tau_i}^{\tau_f} d\tau_1\, d\tau_2 \,e^{- i (E-E_0) (\tau_1-\tau_2) }\, \langle \phi_i | \phi(x_2(\tau_2)) | \phi_f\rangle\langle \phi_f | \phi(x_1(\tau_1)) | \phi_i\rangle. $$ 
 However, it is more interesting to sum over all possible final states of the quantum field $|\phi_f \rangle$.  Thus if we take the initial state $|\phi_i\rangle$ of the field to be its vacuum, using the completeness relation, we have an effective probability of a definitive transition  only in the detector state
\begin{eqnarray}
 {\overline {\cal F}(\Delta E)}= \int {\cal D} \phi_f|{\cal M}_{fi}|^2=g^2 \,|\langle E | \mu | E_0 \rangle|^2 \int_{\tau_i}^{\tau_f} d\tau_1\, d\tau_2 \,e^{- i (E-E_0) (\tau_1-\tau_2) }\, \langle \phi_i | \phi(x_2(\tau_2))\phi(x_1(\tau_1)) | \phi_i\rangle, 
\end{eqnarray}
where
$$\langle \phi_i | \phi(x_2(\tau_2))\,  \phi(x_1(\tau_1)) | \phi_i\rangle \equiv  iG^+(x_2(\tau_2)-x_1(\tau_1)),$$
is the Wightman function. It is then convenient to define two new temporal variables,
$$\tau_+ =\frac{\tau_1+\tau_2}{2}\qquad \Delta \tau = \tau_1-\tau_2.$$
Assuming further adiabatic turn on and off of the detector-field coupling, taking $\tau_i = 0$ and $\tau_f \to \infty$, the response function of the detector per unit $\tau_+$ is defined as (in units of $g^2 \,|\langle E | \mu | E_0 \rangle|^2$ which we shall stick to in the remaining of the paper)
\begin{eqnarray}
 \frac{d {\cal F}(\Delta E)}{ d \tau_+} =  \int_{-\infty }^{\infty} d(\Delta \tau) \, e^{- i \Delta E \Delta \tau } \, i G^{+}(\Delta \tau),
\label{d5}
\end{eqnarray}
where ${\cal F} := \overline{\cal F}/(g |\langle E | \mu | E_0 \rangle|)^2 $ and  we have written $\Delta E = E-E_0$.  $\Delta E >0$ ($\Delta E <0$) denotes excitation (de-excitation) of the detector interacting with the quantum field. The integral \ref{d5} was computed in~\cite{Garbrecht:2004du} in the Bunch-Davies vacuum for a conformal, a massless minimally coupled as well as for a minimally coupled and nearly massless scalar field (see also also~\cite{Birrell:1982ix}). \\

\ref{d5} can be extended to the de Sitter $\alpha$-vacua in  a straightforward manner as follows~\cite{Bousso:2001mw}.  The mode function $u_{\alpha} $ corresponding to the choice of the $\alpha$-vacua  are related to the mode function $u(x)$ corresponding to the Bunch Davies vacuum\footnote{Bunch Davies mode function $u(x)$ assumes a positive frequency appearance at early time i.e. $u(x) \rightarrow e^{-i k \eta + i \vec{k}\cdot\vec{x}}$ for $ k \eta \rightarrow - \infty$ and hence have been the preferred choice of initial state for the quantum fields in de Sitter background.} through a Bogoliubov tranformation~\cite{Allen:1985ux},
\be
u_{\alpha}(x) = \cosh \alpha\, u(x) + \sinh \alpha\, u^{\star} (x), \label{mode}
\ee
where $\alpha$ is a spacetime independent real parameter.  Then the de Sitter invariant Wightman function in these vacua reads~\cite{Allen:1985ux},
\begin{eqnarray}
 G^{+}_{\alpha}(x, x')= \cosh^2 \alpha \,G^{+}(x, x') +\sinh^2\alpha \,G^{+}(\overline{x}, \overline{x}') +\frac12\sinh 2\alpha  \,\left(G^{+}(x, \overline{x}' ) +G^{+}(\overline{x}, x')\right),
\label{d6}
\end{eqnarray}
where a bar over the spacetime points denotes the antipodal position, which corresponds to $\eta \to -\eta$ in \ref{d2}. All the $G^{+}$'s on the right hand side of the above equation stand for the Bunch-Davies vacuum.  For a scalar field of mass $m$ and non-minimal coupling $\xi$, $G^+(x,x')$ reads~\cite{Allen:1985ux},
\begin{eqnarray}
iG^+(x,x') = \frac{H^2}{16\pi^2}\Gamma \left(\frac32 -\nu\right) \Gamma\left(\frac32 +\nu\right)\, _2F_1 \left(\frac32 -\nu, \frac32 +\nu, 2; 1-\frac{y}{4} \right),
\label{d7}
\end{eqnarray}
where 
$$\nu = \left(\frac94 -12 \xi -\frac{m^2}{H^2}\right)^{1/2},$$
and the de Sitter invariant interval $y$ written in terms of the conformal time reads,
$$y(x,x')=\frac{-(\eta- \eta'-i\epsilon)^2+ |\vec{x}-\vec{x}'|^2}{\eta \eta'},$$
where $\epsilon=0^+$. Rewriting things now in the cosmological time $t$ and setting $\vec{x} =\vec{x}'$ for a comoving  detector, we have
\begin{eqnarray}
y(t,t')=-4\left(\sinh\frac{H\Delta t}{2}-i\epsilon\right)^2.
\label{d7''}
\end{eqnarray}
Likewise we have for the antipodal transformations,
\begin{eqnarray}
y(t, \overline{t'})=4 \left(\cosh\frac{H\Delta t}{2}+i\epsilon\right)^2 \qquad y(\overline{t}, t' )=4 \left(\cosh\frac{H\Delta t}{2}-i\epsilon\right)^2 \qquad y(\overline{t}, \overline{t'})= -4\left(\sinh\frac{H\Delta t}{2}+i\epsilon\right)^2.
\label{d7add}
\end{eqnarray}

Since we have set  the comoving spatial separation to zero, the cosmological time $t$ becomes the proper time along detector's trajectory, $\tau=t$. Putting these all in together, the response function  for the de Sitter $\alpha$ vacua is then obtained once we replace the Wightman function appearing in \ref{d5} by \ref{d6},~\cite{Bousso:2001mw}
\begin{eqnarray}
\frac{d {\cal F_{\alpha}}(\Delta E)}{ d t_+} =  \int_{-\infty }^{\infty} d(\Delta t)\, e^{- i \Delta E \Delta t } \, i G_{\alpha}^{+}(\Delta t),
\label{alpha}
\end{eqnarray}
and further use \ref{d7}, \ref{d7''} and \ref{d7add} into it.\\

A few comments pertaining \ref{alpha} are in order here. First,
we note that we have not changed the $e^{- i \Delta E \Delta t }$ term according to the antipodal transformation at all. This is because this term originates purely from the detector, \ref{d4}, and not from the field. Since the detector is a pointlike and localised object, the antipodal transformation simply should not act on this term. 
We also note that
the nonlocal characteristic of the $\alpha$-vacua is manifest from \ref{d6}, where we have added 
Wightman functions corresponding to antipodal points. Thus there seems to be an apparent interpretational problem of coupling a pointlike particle detector to the field in this case. However,  we may get rid of  this issue by recalling that the  Bogoliubov rotation made from the Bunch-Davies modes, \ref{mode}, is purely local.   Since on the other hand the Wightman function does not represent propagation of the field, the same appearing in \ref{alpha} can be interpreted as just an expectation value of the  operator $\phi (x) \phi(x') $ with respect to the vacuum corresponding to the $\alpha$ modes.  With this interpretation, we shall see below that at least for a real scalar field  we can compute the response function in the $\alpha$-vacua without any apparent ambiguity. \\

Finally as discussed before, for a complex scalar field the simplest non-trivial detector-field coupling is quadratic in the field,
\be
{\cal L}_{\rm int}= g \mu (t) \phi^{\dagger}(x(t)) \phi (x(t)). \label{complex}
\ee
Alike the fermions, e.g.~\cite{Louko:2016ptn}, the quadratic coupling is necessary in order to make the field-detector interaction Hamiltonian hermitian. Accordingly, the integral for the response function will contain a product of two Wightman functions. We address this issue in detail in~\ref{s3}.  

\section{Response functions of  a real scalar field in $\alpha$-vacua}\label{free}
\subsection{The conformal scalar }\label{conf}
We start with the simplest case of a conformally invariant scalar field theory. Even though such a field does not create particles in a conformally flat spacetime such as the de Sitter in the conformal vacuum, it is well known that the Unruh-DeWitt detector records a thermal response function in the conformal or the Bunch-Davies vacuum~\cite{Birrell:1982ix}. This result was later extended to the de Sitter $\alpha$-vacua in~\cite{Bousso:2001mw}. We briefly reproduce this result below chiefly to fix the notations we shall be using in the rest of this paper. 

\begin{figure}
	\centering
		\includegraphics[height=6cm]{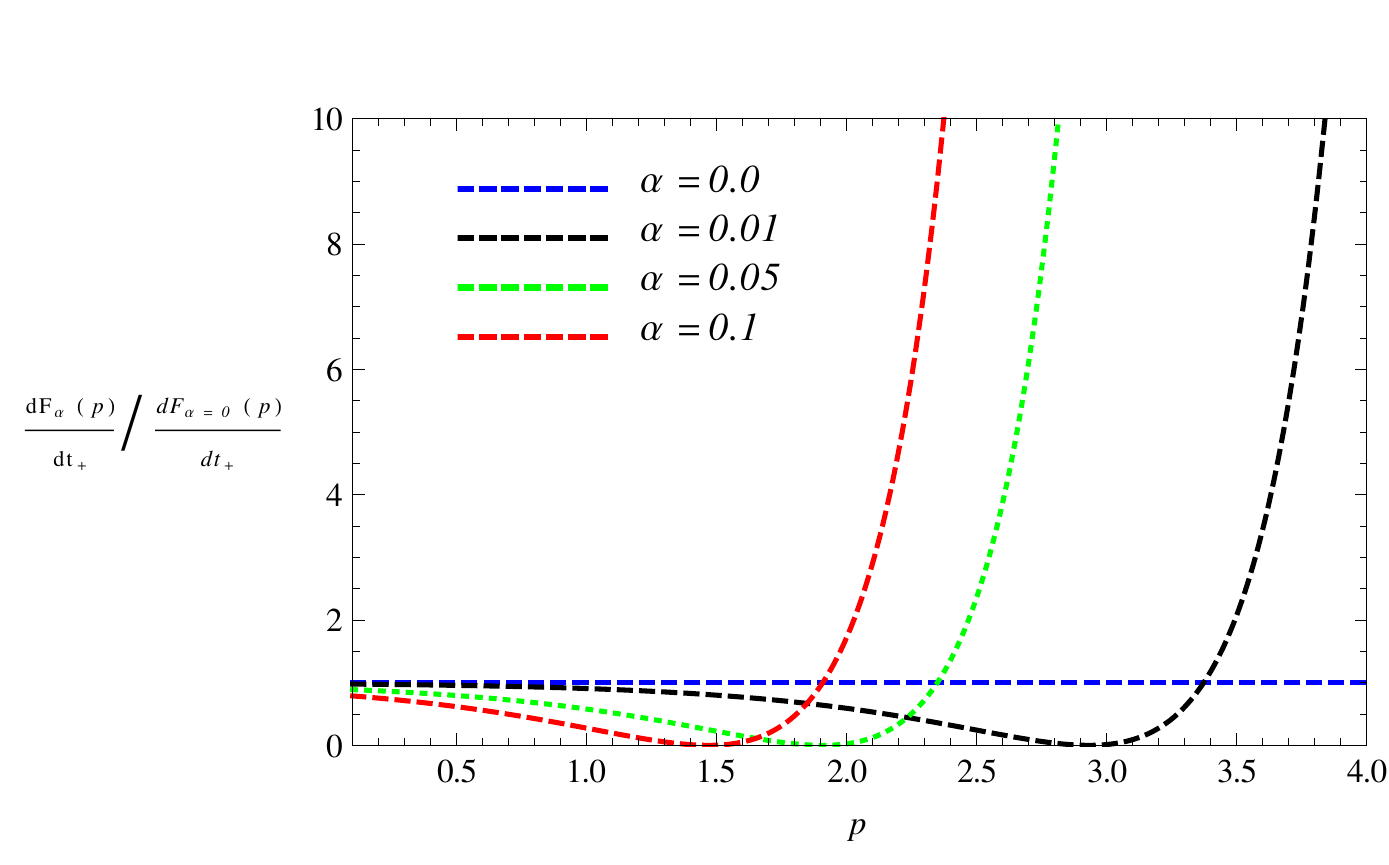}
		\caption{ Variation of the response function \ref{d10}~\cite{Bousso:2001mw}, when scaled by the conformal vacuum ($\alpha=0$) result, with respect to the dimensionless energy difference $p$ for different $\alpha$-values. }
		\label{plot1}
	\end{figure}

In the conformal vacuum ($\nu=1/2$) \ref{d7} in the comoving frame of the detector becomes
\begin{eqnarray}
 iG^{+}(x, x')= \frac{H^2}{4\pi^2 y}= - \frac{H^2}{16\pi^2 (\sinh  \frac{H\Delta t}{2}- i\epsilon)^2},
 \label{d7'}
\end{eqnarray}
and hence \ref{alpha} gives, when written  in terms of a  dimensionless temporal coordinate, $u=H\Delta t/2$,
\begin{eqnarray}
\frac{d {\cal F_{\alpha}}(\Delta E)}{ d t_+} = -\frac{H}{8\pi^2}  \int_{-\infty }^{\infty} d u\,e^{- 2i \Delta E u/H } \left[ \frac{\cosh^2\alpha}{(\sinh  u- i\epsilon)^2} +   \frac{\sinh^2\alpha}{(\sinh u+ i\epsilon)^2}\right. \nonumber\\ \left.-\frac12\sinh 2\alpha\left(\frac{1 }{(\cosh u- i\epsilon)^2}+ \frac{1 }{(\cosh u+i\epsilon)^2} \right)\right].
\label{d9'}
\end{eqnarray}
The above integrals can easily be evaluated e.g. by using a semicircular contour closing in the lower half plane.  Note that since the function $\cosh u$ is never vanishing on the real line, from hereafter we shall not retain the $\pm i \epsilon$ terms for them. We next introduce a dimensionless energy difference for convenience,
$$p:= \frac{2\Delta E}{H},$$
in terms of which the response function is found to be~\cite{Bousso:2001mw},
\begin{eqnarray}
\frac{d {\cal F_{\alpha}}(p)}{ d t_+}\Big\vert_{\rm conf.} = \frac{pH }{4 \pi}\, \frac{\left( \cosh \alpha  - e^{\pi p/2}\sinh \alpha\right)^2}{e^{\pi p }-1 }.
\label{d10}
\end{eqnarray}
Setting $\alpha =0 $ above recovers the thermal spectra associated with the usual conformal vacuum~\cite{Birrell:1982ix}.  We have plotted \ref{d10} scaled by the $\alpha=0$ result in \ref{plot1}. A couple of things are perhaps worth noting : a) the response becomes gradually stronger with the increasing $\alpha$ values b) Taking $\alpha \geq 0$, the numerator of \ref{d10} forces the response function to be vanishing at some $p$-value, for a given $\alpha$ and c) unlike the $\alpha=0$ case~\cite{Birrell:1982ix}, there is no unique meaning of \ref{d10} in terms of pure absorption or pure emission. This should correspond to the fact that firstly, for $\alpha \neq 0$ the timelike parameter corresponding to the Bogoliubov rotated modes, \ref{mode}, is not the proper time along the detector, which dictates its time evolution. Second, such mode mixing makes an $\alpha$-vacuum to be a squeezed state over all the Bunch-Davies states. These two facts force the detector to undergo both excitations and de-excitations, 
as is manifest in \ref{d10}. In different de Sitter vacua the non-thermal response have been obtained from field content analysis as well \cite{Singh:2013pxf}, though an Unruh-DeWitt detector does not always measure  the field content.

	%

\subsection{The minimally coupled massless  scalar }
\label{mmc}
As is evident, one cannot simply set $\nu=3/2$ corresponding to the massless minimal coupling in \ref{d7},  owing to the fact that there exists no de Sitter invariant Wightman function for a massless minimal scalar. One thus needs to find it independently~\cite{Allen:1985ux},
\begin{eqnarray}
iG^{+}(y) = \frac{H^2}{4 \pi^2} \left(\frac{1}{y} -\frac12 \ln y +\frac12 \ln \left(a(\eta) a(\eta') \right) +\ln2 -\frac14 \right).
\label{d9}
\end{eqnarray}
Compared to \ref{d7'}, the above thus contains additional terms including one that breaks the de Sitter symmetry. \ref{alpha} in this case reads, 
\begin{eqnarray}
&&\frac{d {\cal F_{\alpha}}(p)}{ d t_+} =  \frac{H }{2\pi^2}  \int_{-\infty }^{\infty} d u\,e^{- i pu } \left[ -\cosh^2\alpha\left(\frac{1}{4(\sinh  u- i\epsilon)^2} +\frac12\ln \left(-4(\sinh u-i\epsilon)^2\right)\right)\right. \nonumber\\
&&\left.-\sinh^2\alpha \left(\frac{1}{4(\sinh  u+ i\epsilon)^2} +\frac12\ln \left(-4(\sinh u+i\epsilon)^2\right)\right)
 +\sinh2\alpha\left(\frac{1}{2\cosh^2u}  -\ln \left(4\cosh^2 u\right) \right)\right.
\nonumber\\
&&\left. 
+ \left(\frac12 \ln (a(t)a(t'))+\ln2-\frac14\right) e^{2\alpha}    \right],
\label{d9'}
\end{eqnarray}
where $p=2\Delta E/H$ as earlier.  Using  \ref{d10}, we rewrite the above equation as
\begin{eqnarray}
\frac{d {\cal F_{\alpha}}(p)}{ d t_+} =  \frac{d {\cal F_{\alpha}}(p)}{ d t_+}\Big\vert_{\rm conf.} + \frac{H}{2\pi^2}\left(\frac12 \ln (a(t)a(t'))+\ln2-\frac14\right) e^{2\alpha}    \delta(p)
\nonumber\\ 
-\frac{H }{4\pi^2}  \int_{-\infty }^{\infty} d u\,e^{- i pu } \left[ \cosh^2\alpha\, \ln \left(-4(\sinh u-i\epsilon)^2\right)+\sinh^2\alpha\,  \ln \left(-4(\sinh u+i\epsilon)^2\right)
- \sinh2\alpha \ln \left(4\cosh^2 u\right) \right].
\label{d11}
\end{eqnarray}
We note that as $\epsilon \to 0$,
\begin{eqnarray}
{\rm Arg} \left(-4(\sinh u\mp i\epsilon)^2\right) = \pm \pi\, {\rm sgn} (u),
\label{d11'}
\end{eqnarray}
where ${\rm sgn} $ stands for the `sign' function.

Following~\cite{Garbrecht:2004du}, then the regularised form of the logarithmic integrals of \ref{d11} can be found by introducing an infinitesimal positive imaginary part in $p$ and then by integrating them by parts. Some  calculations after using~\ref{d10} yields,
\begin{eqnarray}
\frac{d {\cal F_{\alpha}}(p)}{ d t_+} \Big\vert_{\rm MM}=\frac{p H}{4\pi} \frac{\left( \cosh \alpha  - e^{\pi p/2}\sinh \alpha\right)^2 +(4/p)^2\left( \cosh \alpha  +e^{\pi p /2}\sinh \alpha\right)^2}{e^{\pi p}-1 }\nonumber\\+ \frac{H}{2\pi^2}\left(\frac12 \ln (e^{H(t+t')})+\ln2-\frac14\right) e^{2\alpha}\,    \delta(p) ,
\label{d12}
\end{eqnarray}
where the suffix ``MM'' stands for massless and minimal scalar field. We shall argue in \ref{disc} that the above results for a real conformal or massless minimal scalar goes through arbitrary order of the perturbation theory, without encountering any difficulty. Putting $\alpha=0$  recovers the result of~\cite{Garbrecht:2004du}.

Note that the term proportional to $H(t+t')$ in the above equation is not de Sitter invariant. The physical origin of this is related to the non-existence of a vacuum state for a massless minimal scalar that corresponds to the full isometry of the de Sitter background. This non-existence leads to the de Sitter breaking logarithm in the Wightman function in~\ref{d9}, eventually being reflected in \ref{d12}. In fact such terms are always expected to originate from a state that does not obey the full isometry of the spacetime. For example, even in the flat spacetime one can construct analogous symmetry breaking two point functions for non-vacuum states,  leading to symmetry breaking terms in the response function~\cite{Lochan:2014xja}.

The first term on the right hand side of \ref{d12} diverges as $p \to 0$ whereas the second term diverges at $p=0$.  However, for a detector {\it undergoing}  transitions with {\it discrete}  internal energy levels (i.e., $p= 2\Delta E/H\neq 0$), the $\delta$-function term drops out. On the other hand, for a detector with continuum energy levels, it seems that possibly we need to integrate \ref{d12} over all $p$ values with an appropriate density of states in order  to give the response function a well defined meaning. The behaviour of the first term on the right hand side of \ref{d12} as $p\to 0$ dictates the density of state must vanish at least as ${\cal O}(p^2)$ as $p \to 0$. However, this makes the integral related to the second term vanishing. 

We shall focus in this work {\it only} on detectors  with $p\neq0$, so that we may ignore the $\delta(p)$-term anyway, whenever it appears. The simplest physical example of such a detector is of course, a two level system undergoing level transition.  Even though this makes the de Sitter breaking term in \ref{d12} disappear, for a complex, massless minimal scalar such logarithm will indeed appear from the integration of the Wightman functions, as we shall see below.

We have plotted in \ref{plot2} the characteristics of the response  function ($p\neq 0$), by scaling it with the Bunch-Davies result ($\alpha=0$).  The behaviour is monotonous, as compared to the conformal scalar, \ref{plot1}.
	\begin{figure}
	\centering
		\includegraphics[height=6cm]{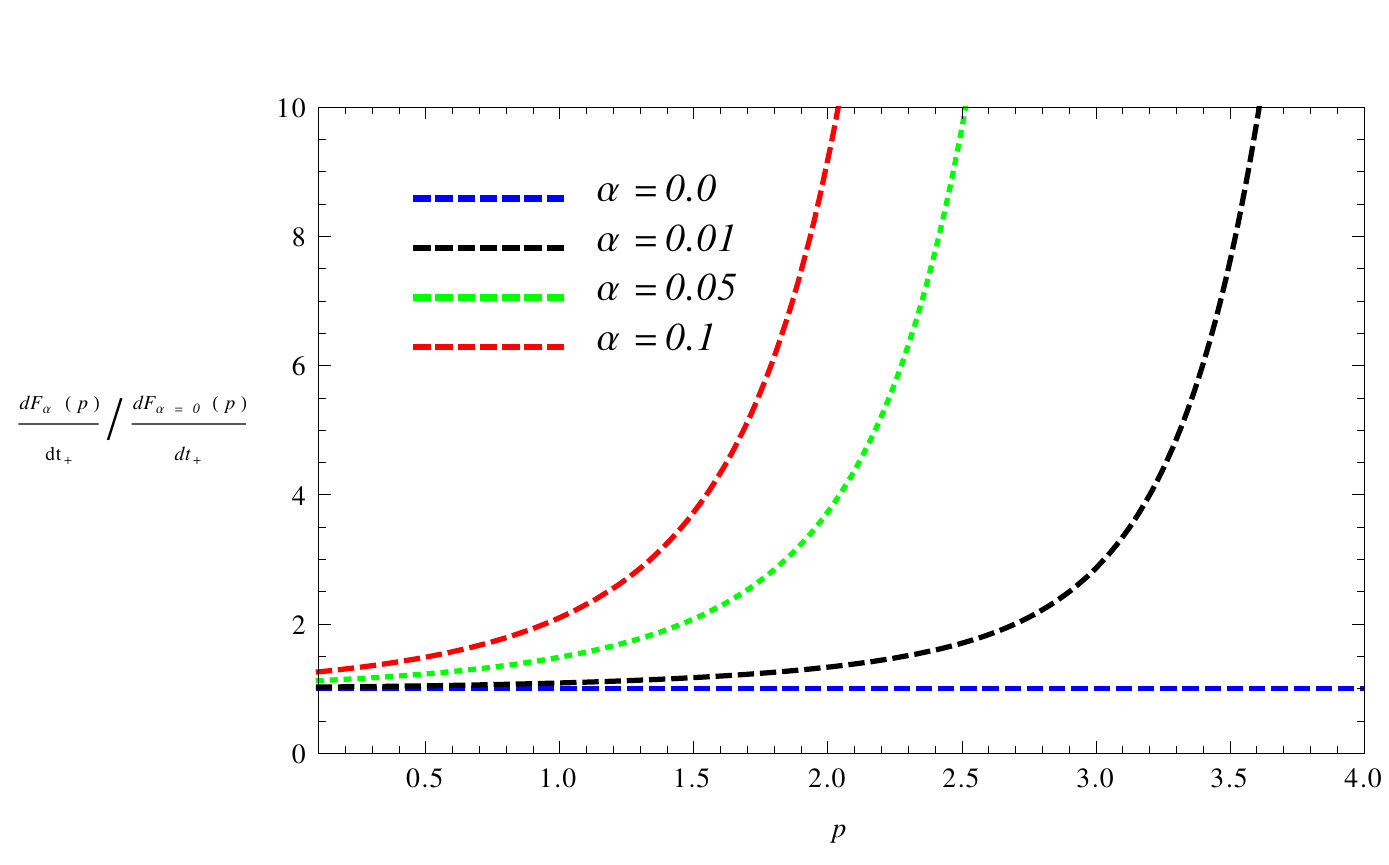}
		\caption{ Variation of the response function \ref{d12} with $p\neq 0$, when scaled by the Bunch-Davies ($\alpha=0$) result, with respect to $p$ for different $\alpha$-values. Unlike \ref{plot1}, the response function is monotonic here. }
		\label{plot2}
	\end{figure}
	%

\section{Complex scalar in the Bunch-Davies vacuum}\label{s3}
\subsection{Conformal complex scalar in conformal vacuum }\label{qcs}
Using \ref{complex}, the first order response function for a complex scalar field is given by,
\begin{eqnarray}
\frac{d {\cal F}(\Delta E)}{ d t_+} =  \int_{-\infty }^{\infty} d(\Delta t) e^{- i \Delta E \Delta t } \left[(i G^{+}(\Delta t))^2 +(i G^+(0))^2\right].
\label{d14}
\end{eqnarray}
The second integrand on the right hand side is divergent and  is present for any spacetime. It was suggested earlier in~\cite{Hinton:1984ht,  Takagi:1984cd} to replace the argument of $iG^+(0)$ by an infinitesimal cut-off and obtain a term proportional to $\delta (p)$,  so that one can  ignore it for transitions with $p \neq 0$. 

A more systematic and satisfactory prescription to deal with this divergence  was  suggested recently in~\cite{Hummer:2015xaa}, by replacing the field-detector interaction Lagrangian density, \ref{complex},  with the normal ordered one,
$${\cal L}_{\rm int}= g \mu (t) \phi^{\dagger}(x(t)) \phi (x(t)) \to g \mu (t) :\phi^{\dagger}(x(t)) \phi (x(t)):$$
Since the $i G^+(0)$ term appearing in \ref{d14} essentially  represents the vacuum to vacuum transition in the coincidence limit, the normal ordering prescription just gets rid of any such divergence beforehand. Thus whenever we have a quadratic field-detector coupling such as \ref{complex}, we must normal order it. Having dealt with this, however, we shall encounter additional divergences (both ultraviolet and infrared) for a massless minimally coupled complex scalar from the first integral of \ref{d14}, which needs further suitable regularization.  
 
However, the first integral does not show any divergence for a conformal scalar  in the conformal vacuum. Using \ref{d7'},  \ref{d14} becomes, 
\begin{eqnarray}
\frac{d {\cal F}(p)}{ d t_+} =  \frac{H^3}{128 \pi^4}\int_{-\infty }^{\infty} d u\,\frac{e^{-   i p u } }{(\sinh u-i\epsilon)^4}.
\label{d15}
\end{eqnarray}
   We can perform the above integration just like the real scalar field by choosing the integration contour to be a semicircle in the lower half plane and taking the poles lying  on the negative imaginary axis. We find 
\begin{eqnarray}
\frac{d {\cal F}(p)}{ d t_+}\Big\vert_{\rm complex,\ conf.} =   \frac{p^3H^3}{384\pi^3}\frac{1}{e^{\pi p}-1}.
\label{d16}
\end{eqnarray}
Note that the thermal factor remains unchanged compared to the real scalar field.

\subsection{The minimally coupled massless complex scalar }\label{qcs2}
Massless complex scalar fields with interactions have previously been studied in context of cosmology \cite{Prokopec:2006ue} and also in BEC systems mimicking  gravity systems \cite{Belenchia:2014hga}, where study of detector response may be more feasible.  Also, as we discussed previously, the quadratic coupling with complex scalars will give us good theoretical exposure of handling the resulting divergences in more physical, e.g. fermionic systems. Therefore, in order to study curvature effects and non-linear couplings in physically realisable systems, we consider complex scalar field interactions as our first step. Further, since we are interested mainly in studying the curvature effects in the detector response, we consider the massless limit first before going to a more realistic small mass limit in  \ref{nmm}.

However, as we shall see this case will not be as simple as that of the conformal one. In fact  there will be finite as well as divergent terms in the expression for the rate of the response function, originating at the coincidence limit of the Wightman functions. We shall not be able to compute the response function in a closed form and will eventually resort to numerical analysis. However, before we do so, we first need to regularise the  integral and also need to cast it into a form which can be handled numerically  without any ambiguity.  \\

We have the  rate of the response function,
\begin{eqnarray}
\frac{d {\cal F}(p)}{ d t_+} =  \frac{H^3}{8\pi^4}\,\int_{-\infty }^{\infty} d u\,e^{- i pu } 
\left[ \frac{(y \ln y -2)^2}{4y^2}  +2\left(\frac12 \ln (a(t)a(t'))+\ln2-\frac14\right)\,\left(\frac{1}{y}-\frac12 \ln y\right)\right. \nonumber\\\left.+ \left(\frac12 \ln (a(t)a(t'))+\ln2-\frac14\right)^2 \right],
\label{d17}
\end{eqnarray}
 which can be rewritten as (after excluding a  $\delta$-function term, cf., the discussion at the end of \ref{mmc}),
\begin{eqnarray}
\frac{d {\cal F}(p)}{ d t_+} =  \frac{H^3}{8\pi^4}\int_{-\infty }^{\infty} d u\,e^{- i p u } 
\left[ \frac{\left[2(\sinh u -i\epsilon)^2 \ln (-4(\sinh u-i\epsilon)^2) +1\right]^2}{16(\sinh u-i\epsilon)^4}  - 
\frac{\left(\frac12 \ln (a(t)a(t'))+\ln2-\frac14\right) }{2 (\sinh u-i\epsilon)^2} \right.\nonumber\\ \left. -\left(\frac12 \ln (a(t)a(t'))+\ln2-\frac14\right)\, \ln (-4(\sinh u-i\epsilon)^2) \right].
\label{d18}
\end{eqnarray}
Let us evaluate the first integral of \ref{d18} first, which is the most non-trivial. It 
reads,
\begin{eqnarray}
\int_{-\infty }^{\infty} d u\,e^{- ip u} \left[\frac{1}{16(\sinh u -i\epsilon)^4}+\frac{\ln \left(-4(\sinh u -i\epsilon \right)^2)}{4(\sinh u -i\epsilon)^2}+\frac{\left(\ln \left(-4(\sinh u -i\epsilon)^2\right) \right)^2}{4}\right].
\label{a4}
\end{eqnarray}
  The first integral in \ref{a4} is the same as that of \ref{d15},
\begin{eqnarray}
\frac{\pi p^3}{48}\,\frac{1}{e^{\pi p}-1}
\label{a0}
\end{eqnarray}
	\begin{figure}[h!]
		\includegraphics[height=6.5cm]{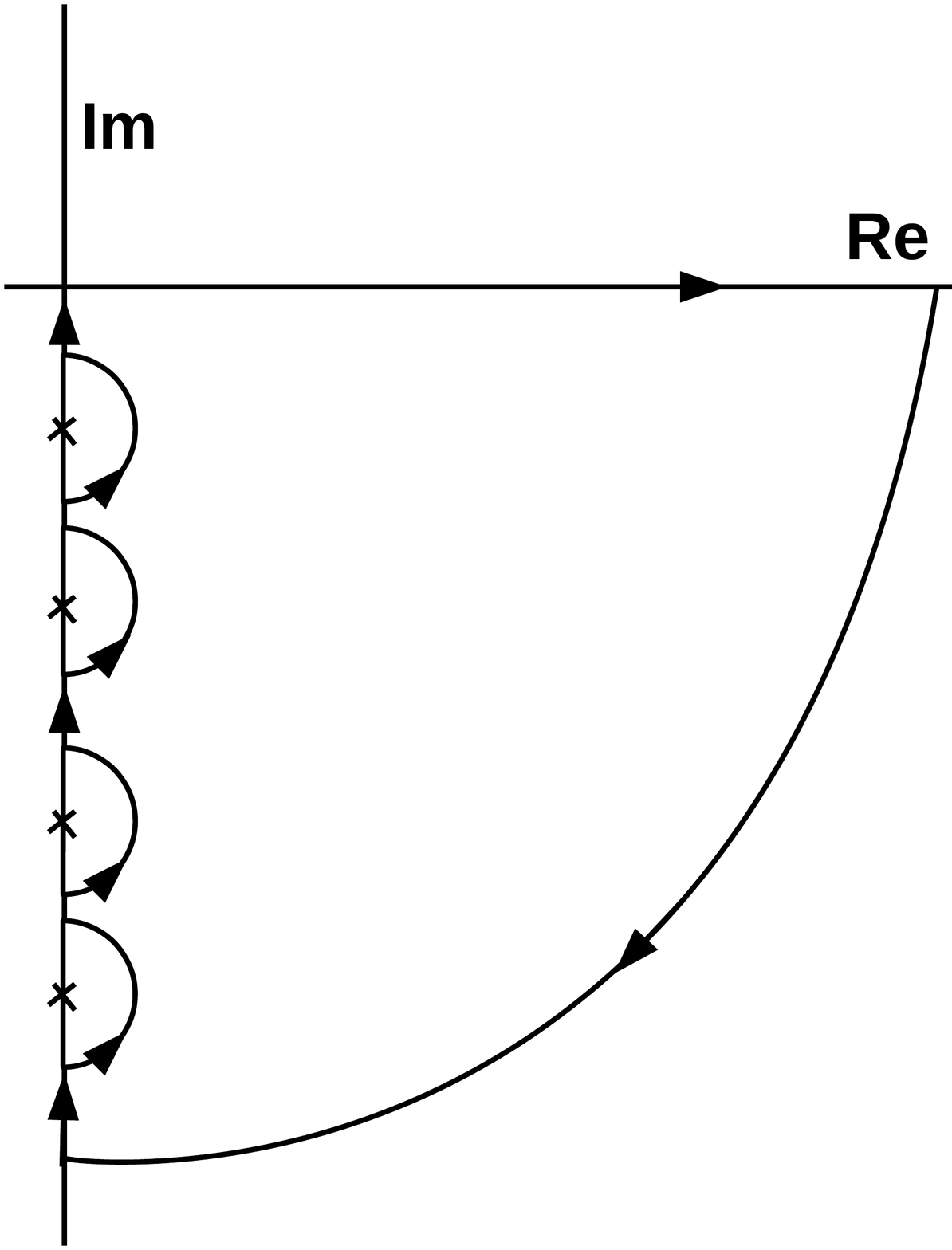}\centering \hskip 2cm
		\includegraphics[height=6cm]{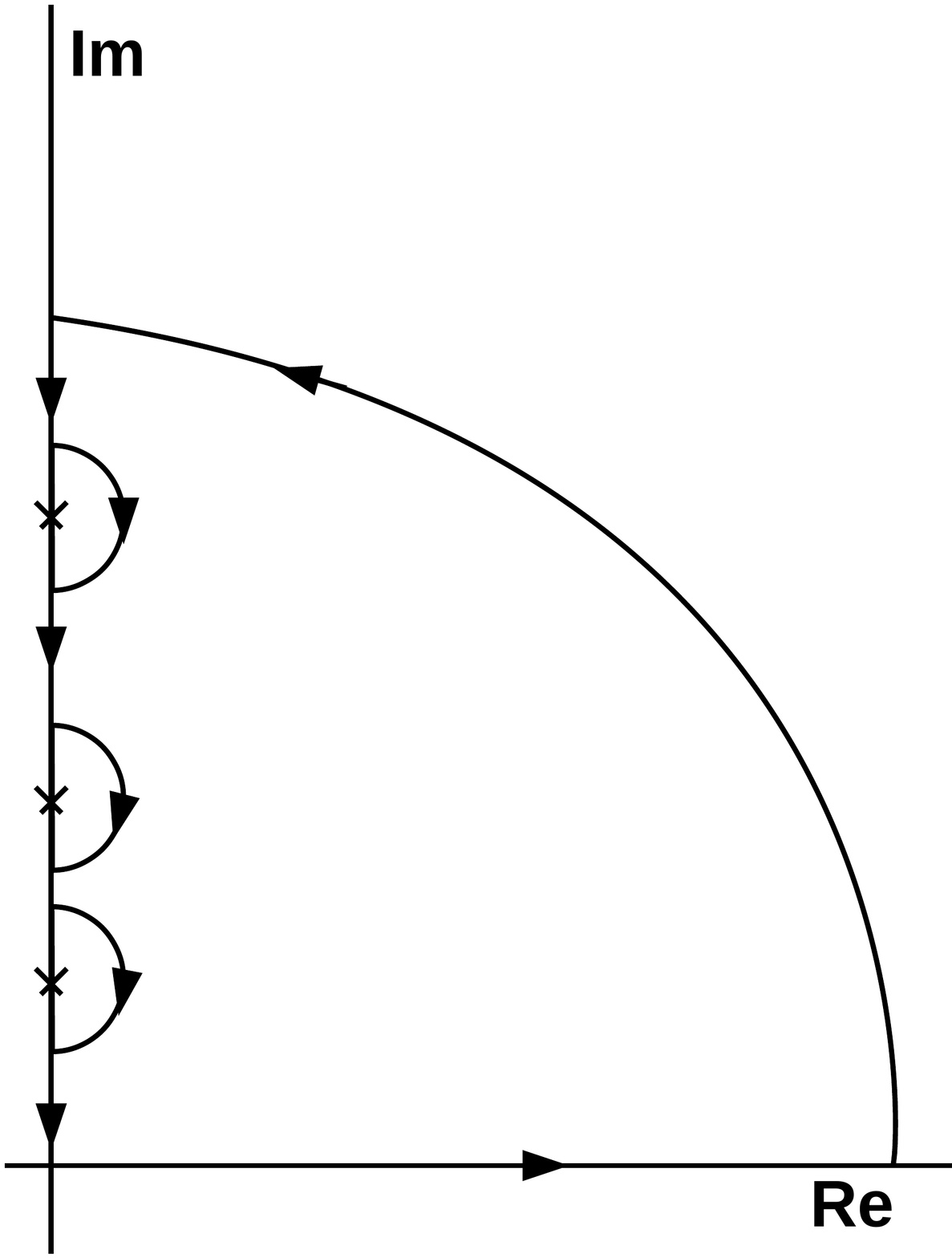}\centering
		\vspace{0.2cm}
		\caption{ Contours to evaluate \ref{n1}.  }
		\label{fig1}
	\end{figure}
Let us now evaluate the second integral of \ref{a4}, which is problematic due to the branch cut of the logarithm. To tackle this, after using \ref{d11'} and expanding the logarithm in powers of $e^{-2u}$, we rewrite it as
\begin{eqnarray}
-\frac12\sum_{n=1}^{\infty}\frac{1}{n} \int_{0}^{\infty} du\, \left(\frac{e^{-(ip +2n)u}}{\left(\sinh u-i\epsilon\right)^2 }\, + \, {\rm c.c.}\right)\,+\, \frac{i}{2}\left(\partial_p+\frac{\pi}{2}\right) \int_{0}^{\infty} du\, \left(\frac{e^{-ipu }}{ \left(\sinh u-i\epsilon\right)^2 }\,-\,{\rm c.c.}\right),
\label{n1}
\end{eqnarray}
where ``c.c." denotes complex conjugation and we have used
\begin{eqnarray}
\int_{-\infty}^{\infty} du\, \frac{e^{-ipu} \, {\rm sgn} (u)}{\left(\sinh u-i\epsilon\right)^2}=\int_{0}^{\infty} \, du \left(\frac{e^{-ipu}}{\left(\sinh u-i\epsilon\right)^2}\, -\, {\rm c.c.}\right).
\label{n4}
\end{eqnarray}
 We shall evaluate the above integral first. We write,
\begin{eqnarray}
\int_{0}^{\infty} \, du \frac{e^{-ipu}}{\left(\sinh u-i\epsilon\right)^2}\, =-i \frac{\partial}{\partial \epsilon}\, \int_{0}^{\infty} \, du \frac{e^{-ipu}}{\left(\sinh u-i\epsilon\right)}.
\label{n5}
\end{eqnarray}
The poles of the integrand are located at
$$u_n= i(n\pi+ (-1)^n \epsilon)\qquad n=0,\,\pm 1,\, \pm 2, \cdots$$
We use a quarter-circular contour in the fourth quadrant  to evaluate this integral, as shown in the first of \ref{fig1} and let the radius of the quarter-circle go to infinity. The poles are avoided using infinitesimal semicircular deformations,
$$z_n=u_n + \overline{\epsilon} e^{i \theta}\quad (\overline{\epsilon} =0^+) \qquad -\pi/2 \leq \theta \leq \pi/2$$
The arc of the quarter-circle does not contribute to the integration. Computing the effect of the deformations and then performing the derivative with respect to $\epsilon$, \ref{n5}, we have 
\begin{eqnarray}
\int_{0}^{\infty} \, du \frac{e^{-ipu}}{\left(\sinh u-i\epsilon\right)^2}\, = -\frac{\pi p}{e^{\pi p}-1}\, +\, i\, \left(\int_{0}^{\infty} du_I \frac{e^{-p u_I}}{(\sin u_I +\epsilon)^2}\right)_{\rm poles~excluded} +{\cal O}(\overline{\epsilon}),
\label{n6}
\end{eqnarray}
where $u_I =\, - {\rm Im} (u)$ along the negative imaginary axis. Note that the poles of the integral on the right hand side  of \ref{n6} are located at  $n\pi- (-1)^n \epsilon,\, n=1,\, 2, \cdots$, which are excluded via the first contour of \ref{fig1}.

As a check of consistency, we have
\begin{eqnarray}
\int_{-\infty}^{\infty} \, du \frac{e^{-ipu}}{\left(\sinh u-i\epsilon\right)^2}\, = \int_{0}^{\infty} \, du \left(\frac{e^{-ipu}}{\left(\sinh u-i\epsilon\right)^2}\, +\, {\rm c.c.}\right)\, =\, -\frac{2\pi p}{e^{\pi p}-1},
\label{n6'}
\end{eqnarray}
where the complex conjugate of the first integral within parenthesis can either be found by using the second contour of \ref{fig1}, or by simply complex conjugating \ref{n6}, since the integral on the right hand side of this equation is real. We can see that \ref{n6'} recovers the result  of \ref{d10} with $\alpha=0$. 

Note that unless we converted the second order pole of the integral to first order via \ref{n5}, the computation of the above infinitesimal deformations of the  contour would not have been possible, for such deformations always yield diverging results for poles beyond first order. \\

We now have from \ref{n4}, \ref{n6},
\begin{eqnarray}
\int_{-\infty}^{\infty} du\, \frac{e^{-ipu} \, {\rm sgn} (u)}{\left(\sinh u-i\epsilon\right)^2}=
2i\,  \left(\int_{0}^{\infty} du_I \frac{e^{-p u_I}}{(\sin u_I +\epsilon)^2}\right)_{\rm poles~excluded}. 
\label{n7}
\end{eqnarray}
The integral on the right hand side is written as, 
\begin{eqnarray}
\int_0^{\pi} du_I \frac{ e^{-p u_I} }{(\sin u_I +\epsilon)^2} + \sum_{n=1}^{\infty} \int_{n\pi + \epsilon}^{(n+1)\pi-\epsilon} du_I \,\csc^2 u_I\, e^{-p u_I}, 
\label{n7-add}
\end{eqnarray}
where $\epsilon = 0^+$ and hence the integration limits exclude the poles. Note that we have got rid of the $\epsilon$ term for the integrals in the summation. Each of the integrals give divergent contribution as we approach the poles.
We tackle with them by treating all the integrals in an equal footing  as follows. We note from \ref{n7} that the coincidence limit of the Wightman function $u\to 0$ on its left hand side corresponds to the points where $\sin u_I = 0$ on the right hand side, achieved via the contour of \ref{fig1}. Thus we shall regularise the integrals of \ref{n7-add} by using suitable regularization near each pole, effectively regularising the very short distance divergent correlation as $u \to 0$ and thereby giving the response function a physical meaning. This task can be largely simplified if we first set all $\epsilon$ terms to zero in \ref{n7-add} and simply rewrite it as
$$\sum_{n=0}^{\infty} \int_{n\pi }^{(n+1)\pi} du_I \,\csc^2 u_I\, e^{-p u_I}, $$
which can be rewritten after a change of variable as,
 $$ \sum_{n=0}^{\infty}\, e^{-p\pi n} \int_{0}^{\pi}\, dx\, \csc^2 x \,e^{-p x}.$$ 
Note that the only poles in the above integral are now located at $x=0,\, \pi$. Performing the sum and further breaking the integration limits, the above can be rewritten as,
$$\cosech \frac{\pi p}{2}\, \int_{0}^{\pi/2} dx  \csc^2 x \, \cosh p\left(\frac{\pi}{2}-x\right), $$
so that the only pole of the above integration is located at $x=0$. Accordingly, we break  the above integration into three pieces and write after using \ref{n7}, 
\begin{eqnarray}
\int_{-\infty}^{\infty} du\, \frac{e^{-ipu} \, {\rm sgn} (u)}{\left(\sinh u-i\epsilon\right)^2} =
2i \left(\coth \frac{\pi p}{2} \int_{0}^{\pi/2} dx\, \csc^2 x\,  -\, p\int_{0}^{\pi/2} dx\, x\,\csc^2 x \right)\,  \nonumber\\+\,2i\cosech \frac{\pi p}{2} \int_{0}^{\pi/2} dx\, \csc^2 x\, \left[\cosh \left(\frac{\pi}{2} -x\right)p - \cosh \frac{\pi p}{2}  +px \sinh \frac{\pi p}{2}\right]. 
\label{divint}
\end{eqnarray}
 The first two integrals in the above expression diverge as $x \rightarrow 0$. It is easy to see from \ref{d17} that there is no flat space limit of this divergence as the integrals vanish in the $H \rightarrow 0$ limit. Such short distance divergence  structure is expected to be present in all Hadamard states.  Before we deal with this divergence, let us also evaluate 
the first integral of \ref{n1}, using the contour of \ref{fig1}. After computing the effect of deformations around the infinitesimal semicircles,  we can express it as,
\begin{eqnarray}
-\frac12\sum_{n=1}^{\infty}\frac{1}{n} \int_{0}^{\infty} du\,\left( \frac{e^{-(ip +2n)u}}{\left(\sinh u-i\epsilon\right)^2 }\, + \, {\rm c.c.}\right) = \frac{\pi p}{e^{p\pi}-1}\sum_{n=1}^{\infty}\frac{1}{n}\,  
+\frac{1}{\sinh \frac{\pi p}{2}}\,\sum_{n=1}^{\infty} \frac{1}{n}\, \int_{0}^{\pi/2}\, dx\, \csc^2x\,\sin 2 n x\, \sinh   \left(  \frac{\pi}{2}-x \right)p,  \nonumber\\
\label{n10}
\end{eqnarray}
The first term on the right hand side corresponds to infinitesimal deformation of the contour in \ref{fig1}, whereas the second integral corresponds to the `poles excluded' part as earlier. For the second integral we slightly lift its lower limit, so  that we can use  the formula 1.441 of~\cite{GR},
\be
\sum_{n=1}^{\infty} \frac{\sin 2n  x}{n}= \frac{\pi -2x}{2}\qquad (0<x<\pi),
\label{sin}
\ee
to rewrite \ref{n10} as
\begin{eqnarray}
\frac{\pi p}{e^{p\pi}-1}\sum_{n=1}^{\infty}\frac{1}{n} + \frac{\pi}{2\sinh \frac{\pi p}{2} } \int_{\epsilon}^{\pi/2}\, dx\, \csc^2 x\, \sinh   \left(  \frac{\pi}{2}-x \right)p - \frac{1}{\sinh \frac{\pi p}{2} } \int_{\epsilon}^{\pi/2}\, dx\, x\, \csc^2 x\, \sinh   \left(  \frac{\pi}{2}-x \right)p.
\label{n10'}
\end{eqnarray}
The first term diverges as $\zeta(1)$ whereas the integrals diverge as $x\to \epsilon$. Accordingly, we now separate the above  into non-divergent  and divergent pieces,
\begin{eqnarray}
&&\frac{\pi p}{e^{p\pi}-1} \sum_{n=1}^{\infty} \frac{1}{n} +\frac{\pi}{2}  \int_{0}^{\pi/2} dx\, \csc^2x - \left(1+\frac{\pi p}{2} \coth \frac{\pi p}{2} \right)\,\int_{0}^{\pi/2} dx\, x \csc^2x \nonumber\\
&&+\frac{\pi}{2\sinh \frac{\pi p}{2} } \int_{0}^{\pi/2}\, dx\, \csc^2 x\, \left[ \sinh   \left(  \frac{\pi}{2}-x \right)p -\sinh \frac{\pi p}{2} +px \cosh \frac{\pi p}{2} \right]\nonumber\\ &&- \frac{1}{\sinh \frac{\pi p}{2} } \int_{0}^{\pi/2}\, dx\, x\, \csc^2 x\, \left[\sinh   \left(  \frac{\pi}{2}-x \right)p-\sinh \frac{\pi p}{2} \right].
\label{n12}
\end{eqnarray}

Combining the above expression and \ref{divint}, we  have the divergent part of \ref{n1} (i.e., the second integral of \ref{a4}),
\begin{eqnarray}
\int_{-\infty }^{\infty} d u\,e^{- ip u}\, \frac{\ln \left(-4(\sinh u -i\epsilon \right)^2)}{4(\sinh u -i\epsilon)^2} \Bigg\vert_{\rm div.} = \frac{\pi}{2} \left(\cosech^2 \frac{\pi p}{2} -\coth \frac{\pi p}{2}+1 \right) \int_{0}^{\pi/2} dx \, \csc^2 x \nonumber\\- \frac{\pi p}{e^{\pi p}-1} \int_{0}^{\pi/2} dx \, x \csc^2 x + \frac{\pi p}{e^{\pi p}-1} \sum_{n=1}^{\infty} \frac{1}{n}
\label{divint2}
\end{eqnarray}
After inserting suitable regulators, we rewrite the second and third of the above divergences as,
\begin{eqnarray}
- \frac{\pi p}{e^{\pi p}-1} \left(\int_{0}^{\pi/2} dx \, x \csc^2 x -  \sum_{n=1}^{\infty} \frac{1}{n}\right) \to - \frac{\pi p}{e^{\pi p}-1} \left(\int_{0}^{\pi/2} dx \, x \csc^2 (x+\epsilon) -  \sum_{n=1}^{\infty} \frac{(1-\epsilon')^n}{n}\right)= \frac{\pi p}{e^{\pi p}-1} \ln \frac{\epsilon}{\epsilon'} \nonumber\\
\label{divint3}
\end{eqnarray}

Note that we could have set $\epsilon = \epsilon'$ above to get rid of the term as a whole. However, since these cut-offs are used to regularise respectively, an integration and a series, for the sake of generality we have kept them distinct. 

Likewise the first divergence of \ref{divint2} can be regularised by setting $\csc x \to \csc (x+\epsilon)$, and we have
\begin{eqnarray}
\int_{-\infty }^{\infty} d u\,e^{- ip u}\, \frac{\ln \left(-4(\sinh u -i\epsilon \right)^2)}{4(\sinh u -i\epsilon)^2} \Bigg\vert_{\rm div.} = \frac{\pi \coth\frac{ \pi p}{2}}{e^{\pi p}-1}\frac{1}{\epsilon}+\frac{\pi p}{e^{\pi p}-1} \ln \frac{\epsilon}{\epsilon'}
\label{divint3'}
\end{eqnarray}
The right hand side is divergent in the absence of regulators introduced and therefore depends upon the fashion in which the regularisation  is carried out. Interestingly these divergences can easily be tackled through the renormalisation of the off-diagonal matrix elements of the detector's monopole operator, $\mu$, in its energy eigenbasis, as follows\footnote{This result can also be obtained through the identification $\epsilon =\epsilon'$, followed by a principal value computation of the first integral on the right hand side of \ref{divint2}. There might exist some other viable regularisation or renormalisation schemes as well to tackle this divergence. Of course, like any other divergent observable in QFT, the physically meaningful finite result may be different in different regularisation schemes and only experiments can ascertain which scheme is correct.}. We modify the field-detector interaction coupling by adding another monopole operator that does {\it not} couple to the field,
$$ {\cal L}_{\rm int}=  g \mu (t) :\phi^{\dagger}(x(t)) \phi (x(t)): +  \mu'(t)$$
Analogous modification for a real scalar field yields the shift of the detector energy level, as has been discussed recently in~\cite{Kaplanek:2019dqu}. The response function with the above modification becomes (after ignoring a $\delta$-function  as earlier),
\begin{eqnarray}
\frac{d {\cal F}(p)}{ d t_+} =  \frac{2}{H}\int_{-\infty }^{\infty} du \,e^{- i pu } \,(i G^{+}(u))^2 + \frac{2}{gH} \left( \frac{\langle E| \mu' | E_0\rangle}{\langle E| \mu | E_0\rangle}+  {\rm c.c.}\right)\int_{-\infty }^{\infty} du \,e^{- i pu } \,i G^{+}(u)
\label{divint4}
\end{eqnarray}

The first term on the right hand side  gives the usual response function integral for a massless minimal complex scalar, \ref{d18}, whereas the second term yields the response function for a massless minimal {\it real} scalar, \ref{d12} (with $\alpha =0$ and $p\neq 0$). Our objective is to exploit this new term to cancel the contribution \ref{divint3'}. We find the desired condition after a little calculation,
\begin{eqnarray}
\langle E |\mu'|E_0 \rangle = -\frac{g H^3 }{8\pi^2 p \left(1+16/p^2 \right)}\left( \frac{\coth \frac{\pi p}{2}}{\epsilon} + p \ln \frac{\epsilon}{\epsilon'}\right) \langle E |\mu|E_0 \rangle
\label{divint5}
\end{eqnarray}
This  corresponds to an operator relationship, 
\begin{eqnarray}
\mu' = - \frac{g H^3}{8\pi^2 } \, \sum_{i,j,\, i\neq j } C_{ij}  | i \rangle \langle i | \mu | j \rangle \langle j |
\label{divint6}
\end{eqnarray}
where the states  represent  the complete orthonormal  energy eigenkets of the detector, and $C_{ij}$'s are real numbers depend upon the energy levels. Denoting then $|E_0\rangle  $ and $| E\rangle$ respectively by, for example $| i=0\rangle  $ and $| i=1\rangle$, we obtain from \ref{divint5}, \ref{divint6}, the requirement
\be
C_{10}(p=2(E-E_0)/H ) = \frac{1}{p (1+16/p^2)}\left(\frac{\coth \frac{\pi p}{2} }{\epsilon} + p \ln \frac{\epsilon}{\epsilon'}\right)
\label{A1}
\ee
It is clear that by construction, $\mu'$ will cancel the divergence for any level transition  of the detector. Note also that $C_{10}$ is even in $p$, i.e. $C_{01}= C_{10}$. Likewise $C_{ij} =C_{ji}$ for all $i,j$.

Note that $\mu'$ is non-diagonal in the energy eigenbasis of the detector, \ref{divint6}. 
Thus it is clear from \ref{divint5} that here we are basically renormalising the off-diagonal matrix elements of the operator $\mu$ via the  operator $\mu'$, which serves as a  counterterm. However in doing so,   we have the appearance of a detector-curvature interaction, $gH^3$, in \ref{divint6}. This seems to be curious, for such non-minimal interaction of the detector and the spacetime curvature is introduced at the level of the Lagrangian density itself. It will be interesting to see whether this term can produce any finite or observable effects at  higher order of the perturbation theory. We hope to pursue this issue  in a  future work.  

We also refer our reader to~\cite{Hummer:2015xaa} for  a discussion on the renormalisation of the Unruh-DeWitt detector response in the flat spacetime with different  interacting field theories like QED. \\

 Collecting now the finite pieces from \ref{divint}, \ref{n12}  and using  \ref{n1}, we obtain the regularised expression of the second integral of \ref{a4}, 
\begin{eqnarray}
&&\int_{-\infty }^{\infty} d u\,e^{- ip u}\, \frac{\ln \left(-4(\sinh u -i\epsilon \right)^2)}{4(\sinh u -i\epsilon)^2} \Bigg\vert_{\rm Regularised} \nonumber\\=&&\frac{\pi}{2} \cosech \frac{\pi p}{2} \coth\frac{\pi p}{2} \, \int_{0}^{\pi/2}\, dx\, \csc^2 x\, \left[ \cosh   \left(  \frac{\pi}{2}-x \right)p -\cosh \frac{\pi p}{2} + px \sinh \frac{\pi p}{2}\right] \nonumber\\&&-\cosech \frac{\pi p}{2} \, \int_{0}^{\pi/2}\, dx\, \csc^2 x\, \left[ \left(\frac{\pi}{2}-x \right) \sinh \left(\frac{\pi}{2}-x \right)p -   \left(\frac{\pi}{2}-x \right) \sinh \frac{\pi p}{2} + \frac{\pi p x}{2} \cosh \frac{\pi p}{2}\right]
 \nonumber\\ &&- \frac{\pi}{e^{\pi p}-1}\, \int_{0}^{\pi/2}\, dx\, \csc^2 x\, \left[e^{px}-1-px \right] - \cosech \frac{\pi p}{2} \, \int_{0}^{\pi/2}\, dx\, x\,\csc^2 x\, \left[ \sinh   \left(  \frac{\pi}{2}-x \right)p -\sinh \frac{\pi p}{2}\right].\nonumber\\
\label{second} 
\end{eqnarray}

Finally, we come to the third integral of \ref{a4}. 
After using \ref{d11'}, it takes the form (after ignoring a term containing $\delta (p)$),
\begin{eqnarray}
\frac12 \int_{0 }^{\infty} d u\, \cos p u \left(\ln (4\sinh^2 u)\right)^2 +\pi \int_{0}^{\infty} du\, \sin {p u}\, \ln (4\sinh^2 u). 
 \label{a5}
\end{eqnarray}
After expanding the logarithms, the above integral can be rewritten as, 
\begin{eqnarray}
&&-2 \left(\partial_p^2 + \pi \partial_p\right)\, \int_{0}^{\infty} du\, \cos pu - 4 \sum_{n=1}^{\infty} \frac{1}{n} \,\partial_p \int_{0}^{\infty} du\, \sin pu \,e^{-2nu} \nonumber\\ &&-2\pi \sum_{n=1}^{\infty} \frac{1}{n} \int_{0}^{\infty} du\, \sin pu\, e^{-2nu} + 2 \sum_{m,n=1}^{\infty} \frac{1}{mn} \int_{0}^{\infty} du\, \cos pu \,e^{-2(m+n)u}.
 \label{n14}
\end{eqnarray}

The first two integrals diverge as $u \to \infty$. Such infrared divergence can be regularised by introducing an
infinitesimal positive imaginary part in $p$. Accordingly, we get
\begin{eqnarray}
 \int_{0}^{\infty} d u\,\cos pu = 0.
 \label{a2}
\end{eqnarray}
Using this
and also integrating by parts, \ref{n14} can be put into a regularised form
\begin{eqnarray}
 4 \sum_{n=1}^{\infty} \frac{p^2 -4n^2}{n(p^2 +4n^2)^2} - \sum_{n=1}^{\infty} \frac{2\pi p}{n(p^2+4n^2)}+ 8 \sum_{m,n=1}^{\infty}\, \frac{1}{n\, (p^2 + 4(m+n)^2)}.
\label{third}
\end{eqnarray}
The various summations appearing above, as is evident, are all convergent. Thus  \ref{a0}, \ref{second} and \ref{third}, when added together, gives a regularised expression  for the first integral of \ref{d18}. 

The remaining integrals of \ref{d18}, i.e. the second and the third ones were already evaluated respectively in \ref{conf} and \ref{mmc}. Thus, combing them with \ref{a0}, \ref{second} and  \ref{third}, we finally obtain a fully regularised expression of the detector response function of \ref{d18}, 
\begin{eqnarray}
&&\frac{d {\cal F}(p)}{ d t_+}\Big \vert_{\rm complex,\, MM} =\frac{p^3H^3}{384 \pi^3}\,\frac{1}{e^{\pi p}-1}+\left(\frac12 \ln (a(t)a(t'))+\ln2-\frac14\right)\,\frac{p H^3}{8\pi^3}\, \frac{1}{e^{\pi p}-1} \left(1 + \frac{8}{p^2}\right)
\nonumber\\
&&+\frac{H^3}{2\pi^4} \left[ \sum_{n=1}^{\infty} \frac{p^2 -4n^2}{n(p^2 +4n^2)^2} - \sum_{n=1}^{\infty} \frac{\pi p}{2n(p^2+4n^2)}+  2\sum_{m,n=1}^{\infty}\, \frac{1}{n\, (p^2 + 4(m+n)^2)}\right] \nonumber\\
&&+\frac{H^3}{16\pi^3} \cosech \frac{\pi p}{2} \coth \frac{\pi p}{2} \, \int_{0}^{\pi/2}\, dx\, \csc^2 x\, \left[ \cosh   \left(  \frac{\pi}{2}-x \right)p -\cosh \frac{\pi p}{2} + px \sinh \frac{\pi p}{2}\right] \nonumber\\&&-\frac{H^3}{8\pi^4}\cosech \frac{\pi p}{2} \, \int_{0}^{\pi/2}\, dx\, \csc^2 x\, \left[ \left(\frac{\pi}{2}-x \right) \sinh \left(\frac{\pi}{2}-x \right)p -   \left(\frac{\pi}{2}-x \right) \sinh \frac{\pi p}{2} + \frac{\pi p x}{2} \cosh \frac{\pi p}{2}\right]
 \nonumber\\ && - \frac{ H^3}{8\pi^4}\cosech \frac{\pi p}{2} \, \int_{0}^{\pi/2}\, dx\, x\,\csc^2 x\, \left[ \sinh   \left(  \frac{\pi}{2}-x \right)p -\sinh \frac{\pi p}{2}\right] - \frac{H^3}{8\pi^3}\frac{\pi}{e^{\pi p}-1}\, \int_{0}^{\pi/2}\, dx\, \csc^2 x\, \left[e^{px}-1-px \right]. \nonumber\\
\label{final}
\end{eqnarray}

Before we proceed, we note the  regularization procedures we adopted to derive the above expression. First, we discussed a couple of possible regularization cum renormalization schemes to tackle the short distance ultraviolet divergence of \ref{divint2}. Second, we also needed to introduce an infinitesimal positive imaginary part in $p$, in order to regularise the infrared divergence of some of the integrals of \ref{n14}.

If we let $p \to \infty$ in \ref{final}, each of the terms of the first two lines as well as the last integral vanish. The remaining three integrals do not vanish individually in this limit, but they cancel with each other to yield a vanishing contribution. This is expected, as $p \to \infty$ corresponds to energy level separation of the detector much larger than the spacetime energy scale, $H$.
 
Note that there is a logarithm  in \ref{final}, increasing monotonically with 
$2H t_+= H(t+t')$, indicating breakdown of the perturbation theory as $Ht, Ht' \gg1 $. This seems to be analogous to the secular growth reported  earlier in the context of perturbative quantum field theory in de Sitter space, e.g.~\cite{Brunier:2004sb, Karakaya:2019vwg, Baumgart:2019clc, Woodard:2014jba, Moreau:2018lmz} (also references therein). Such secular growth is absent for a real massless minimal scalar, \ref{d12}, for in that case the de Sitter breaking term is accompanied by a $\delta$-function. We also note that in this limit the $p$-dependence of \ref{final}  becomes qualitatively similar to that  of the real scalar,~\ref{d12} (with $\alpha=0$ and $p \neq 0$). Finally, we note that \ref{final} diverges for small $p$-values, as of the real scalar, \ref{d12}.

\begin{figure}
	\centering
		\includegraphics[height=6cm]{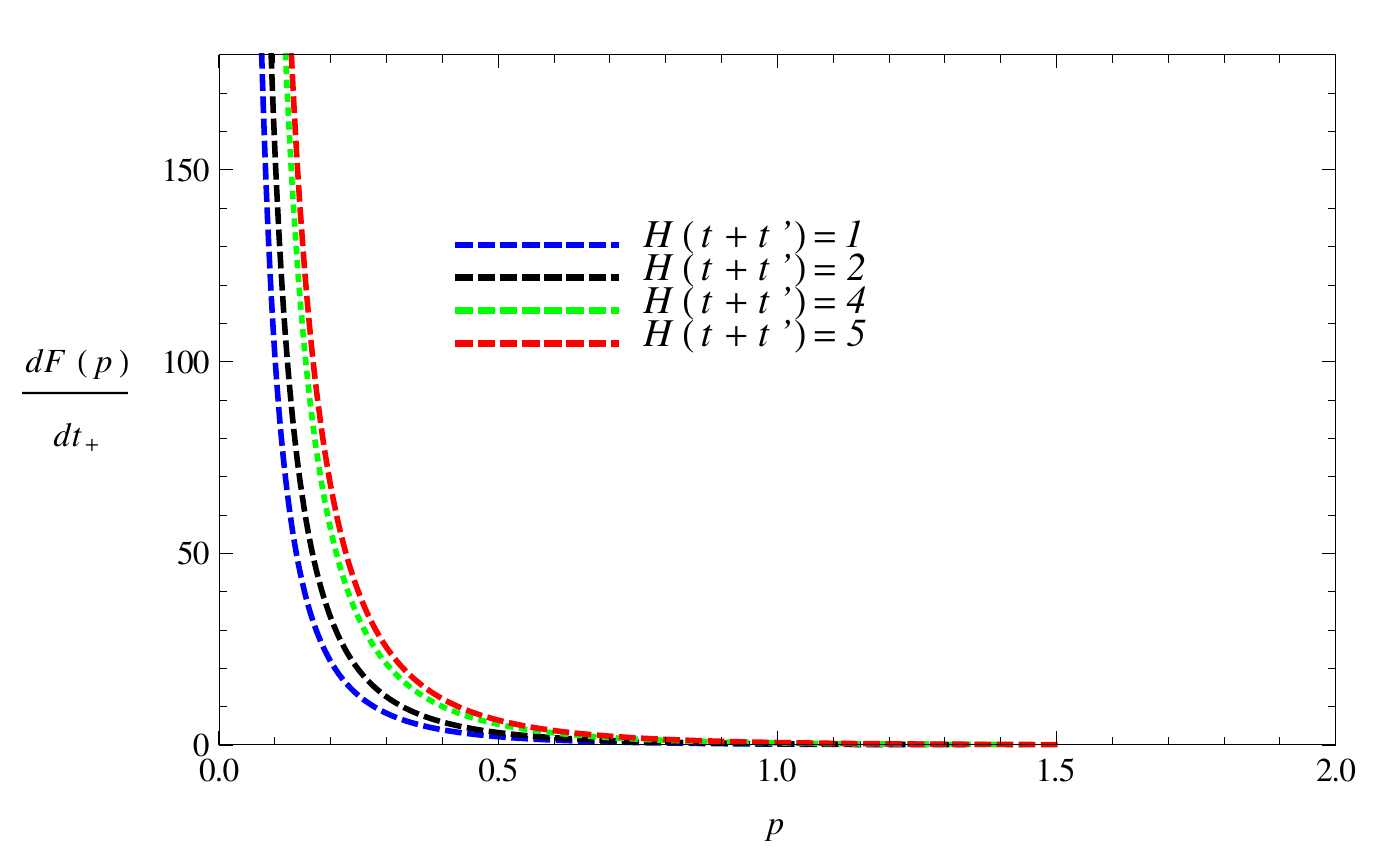}
		\caption{ Plot of \ref{final} (after scaling it by $H^3/8\pi^3$) with respect to $p$ for different values of $H(t+t')=2Ht_+$. }
		\label{plot3}
	\end{figure}

Since the response function \ref{final} is now regularised,  we  investigate its behaviour numerically without any trouble, as a function of the dimensionless energy $p$, \ref{plot3}.  
	%

\section{The ambiguity of complex scalar field with $\alpha$-vacua}\label{alpha-com}

We finally come to the case of complex scalar fields with $\alpha$-vacua. However, we  argue below that the detector response function is not well defined in this case. The response function for a conformal scalar in this case is given as,
\begin{eqnarray}
\frac{d {\cal F_{\alpha}}(p)}{ d t_+} =  \frac{H^3}{128 \pi^4}\, \int_{-\infty }^{\infty} d u\,e^{- i p  u } \left[ \frac{\cosh^2\alpha}{(\sinh  u- i\epsilon)^2} +   \frac{\sinh^2\alpha}{(\sinh u+ i\epsilon)^2}-\frac{\sinh 2\alpha}{\cosh^2 u} \right]^2.
\label{d20}
\end{eqnarray}
Expanding the square, making some rearrangements of terms and also  redefining $\epsilon$ in some of the integrals, we find
\begin{eqnarray}
\frac{d {\cal F_{\alpha}}(p)}{ d t_+} =  \frac{H^3}{128 \pi^4}\, \int_{-\infty }^{\infty} d u\,e^{- i p u} \left[ \frac{\cosh^4\alpha}{(\sinh  u- i\epsilon)^4} +  \frac{\sinh^4\alpha}{(\sinh  u+ i\epsilon)^4}
+\frac{\sinh^2 2\alpha}{2}\left(\frac{1}{\sinh^4u}+\frac{2}{\cosh^4u}\right)\right. \nonumber\\ \left. 
- 8\sinh2\alpha \left(\frac{\cosh^2\alpha}{(\sinh 2u -i\epsilon)^2}+\frac{ \sinh^2 \alpha}{(\sinh 2u +i\epsilon)^2}\right)\right].
\label{d21}
\end{eqnarray}
Note that there is an integral  containing  $\sinh^{-4}u$ without any $i\epsilon$. This  term arises due to the multiplication  
$(\sinh u -i\epsilon)^2(\sinh u +i\epsilon)^2$,  while squaring. All but the integral containing this term can be straightforwardly evaluated.  The integral containing $\sinh^{-4}u$ is problematic because it does not converge on the real line, nor we can attempt to compute any principal value, for it diverges in the presence of poles beyond the first order. We cannot re-insert any $i\epsilon$ term now, for the answer will depend upon the sign of that term.  Also, the contour of \ref{fig1} cannot be used here, for we can compute the effect of the infinitesimal semicircular deformations to avoid the poles only if the poles are of first order. 


Similar cancellation of the $i\epsilon$ regulator and some other ambiguities in the context of the perturbation theory in de Sitter $\alpha$-vacua  were reported earlier in~\cite{Einhorn:2002nu, Banks:2002nv, Collins:2003zv, Einhorn:2003xb, Collins:2003mj, Collins:2004wi, Collins:2004wj, Collins:2007jc, Collins:2008xk}. 
Such ambiguity seems to originate from the  inherent non-local characteristic of the $\alpha$-vacua, coming from the antipodal transformations discussed in \ref{setup}. Thus even though we may give the detector response function for a real scalar in the $\alpha$-vacua a meaning in the sense of just an expectation value (cf. the discussions at the end of \ref{setup}), apparently it fails for a complex scalar.  
In~\cite{Collins:2004wj} (also references therein), it was suggested  to modify the Feynman propagator by adding two sources,  in order to tackle the  non-locality. However, for a pointlike, localised  particle detector it is not clear to us how much appropriate will be any such analogous modification.  Since such a detector model is very generic and well motivated and also keeping in mind the   existing results on the de Sitter $\alpha$-vacua mentioned above,  to the best of our knowledge and understanding, it seems that the above problem should  not  be attributed to the model of the detector we are working in, but it should be regarded  as a generic problem of the perturbation theory with the $\alpha$-vacua.\\

\begin{figure}[h!]
	\includegraphics[height=8cm]{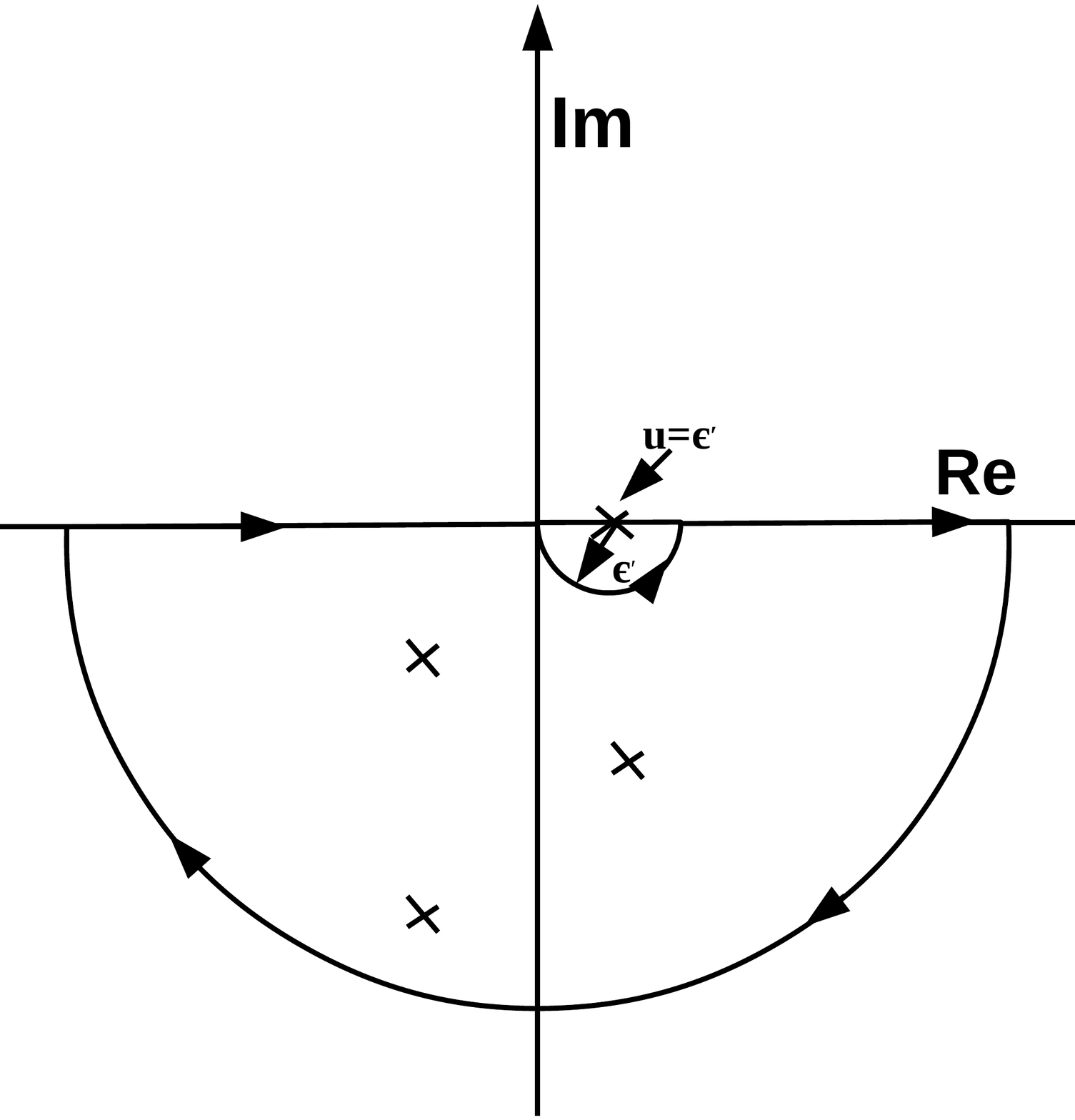}\centering \hskip 2cm
		\includegraphics[height=8cm]{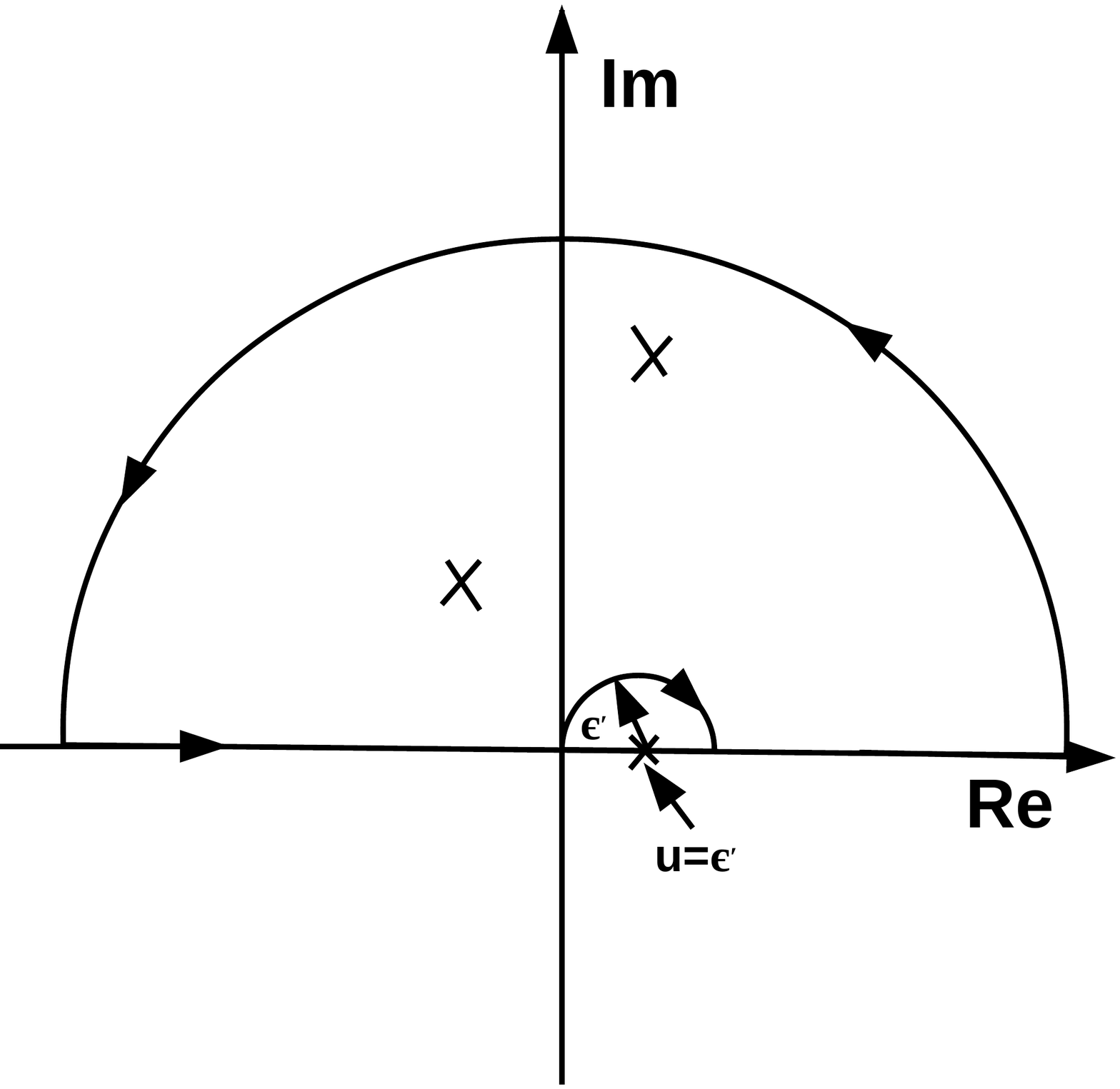}\centering
		\vspace{0.2cm}
		\caption{ Possible contours to evaluate \ref{d22}. }
		\label{fig2}
	\end{figure}

Nevertheless, we may still try to obtain a regularised version of the problematic integral as follows. However the caveat is, this will {\it not} yield  a unique result, as described below.  Let us  first rewrite the integral as 
$$\int_{-\infty}^{\infty}du\, \frac{e^{-ipu}}{\sinh^4u} = \frac{1}{12} \lim_{\epsilon' \to 0} \partial_{\epsilon'}^3 \left(\int_{-\infty}^{\infty} du \, \frac{e^{-ipu}}{(\sinh u -\epsilon')} +{\rm c.c.} \right),$$
where $\epsilon'$ is  real, no matter positive or negative.
Since the pole on the real axis of the integrand  on the right hand side is of first order, its principal value is well defined. This allows us  to thus {\it define} a  regularised value of our original integration as,
\begin{eqnarray}
\int_{-\infty}^{\infty}du\, \frac{e^{-ipu}}{\sinh^4u}\Bigg \vert_{\rm Regularised} := \frac{1}{12} \lim_{\epsilon' \to 0} \partial_{\epsilon'}^3\, \left( {\rm PV}\,\int_{-\infty}^{\infty} du \, \frac{e^{-ipu}}{(\sinh u -\epsilon') }+ {\rm PV}\,\int_{-\infty}^{\infty} du \, \frac{e^{ipu}}{(\sinh u -\epsilon') }\right).
\label{d22}
\end{eqnarray}
 Accepting this definition, the contour for the first  integration on the right hand side is taken to be a semicircle with an infinitesimal semicircular deformation of radius $\epsilon'$ centred at $u=\epsilon'$,  in the lower half plane. Thus the poles we pick up are located at $u_n=  in\pi+ (-1)^n\epsilon'$ ($n=-1, -2, -3, \dots$). For the second, we use similar contour in the upper half plane, picking up the poles at  $u_n=  in\pi+ (-1)^n\epsilon'$ ($n=1, 2, 3, \dots$), \ref{fig2}. Splitting the integral into two parts on the right hand side in the definition of \ref{d22} will ensures its real valuedness. 
	We find
\begin{eqnarray}
\int_{-\infty}^{\infty}du\, \frac{e^{-ipu}}{\sinh^4u}\Bigg \vert_{\rm Regularised}
= -\frac{\pi p^3}{3} \,\left[ \frac{1}{e^{\pi p}+1} -\frac{1}{2 }+\frac{2}{p^2}\right]. \label{d24}
\end{eqnarray}

Evaluating the rest of the integrals in \ref{d21}, we obtain a regularised detector response function for a conformal complex scalar in the $\alpha$-vacua,
\begin{eqnarray}
\frac{d {\cal F_{\alpha}}(p)}{ d t_+}\Big \vert_{\rm complex\, conf.}=\frac{p^3 H^3}{384 \pi^3}\left[ \frac{\cosh^4 \alpha +\sinh^4 \alpha \,\,e^{\pi p}}{e^{\pi p}-1}  +  \frac{12 \sinh 2\alpha \left(\cosh^2 \alpha +\sinh^2 \alpha\, e^{\pi p/2}\right)}{p^2 (e^{\pi p/2}-1)}\right. \nonumber\\ \left.
-\frac12 \sinh^2 2\alpha \,\left( \frac{e^{\pi p}-1-2 e^{\pi p/2}(e^{\pi p}+1)  } {e^{2\pi p}-1} -\frac12 +\frac{2}{p^2} \right)  \right].\label{d25}
\end{eqnarray}
Setting $\alpha=0$ recovers the result of \ref{d16}.

As  evident, the above result is not unique, for \ref{d24} would change if we change the integration contour. For example, in the first of \ref{fig2}, we could have made the infinitesimal semicircular deformation in the upper half plane as well (and the opposite in the second of \ref{fig2}), which will lead to a change of sign of the second and third terms appearing on the right hand side of  \ref{d25}. We face similar ambiguity for the case of a massless minimal scalar field as well. Again, such ambiguities can be attributed towards regularisation scheme adopted for the divergent integrals, and in the absence of any limiting consistency condition for the $\alpha$-vacua, perhaps only experiments should be able to verify or rule-out the correct regularisation for obtaining physical finite results.

\section{The case of a nearly minimally coupled massless scalar }\label{nmm}
We shall end with a comment on the case of a  nearly massless minimally coupled scalar. The Wightman function in this case reads,
\begin{eqnarray}
iG^{+}(y) = \frac{H^2}{4 \pi^2} \left(\frac{1}{y} -\frac12 \ln y +\frac{1}{2s} +\ln2 -1 \right),
\label{nms1}
\end{eqnarray}
where $s=3/2-\nu$ with $|s|\ll 1$ is a small parameter and $y$ is given by \ref{d7''}. Note that the de Sitter symmetry breaking logarithm is absent here compared to~\ref{d11}. Comparing \ref{nms1} and \ref{d11} it is clear that we can compute the detector response for this scalar by just making the replacement,
$$\left(\frac12 \ln (a(t)a(t'))+\ln2-\frac14\right)\, \to\,\left(\frac{1}{2s} +\ln2 -1\right),$$
appearing in any of the expressions for the response function for the massless minimal scalar field (such as \ref{final}). Thus  despite $s$ is large, there will be no term growing with time in this case, as compared to the exactly massless one. On the other hand, if we take a complex scalar field in the $\alpha$-vacua, problems exactly similar to \ref{alpha-com} will prevail.

\section{Discussions} \label{disc}
We have computed, in this work, the response function for the Unruh-DeWitt detector  coupled to a complex scalar field at the first order perturbation theory, for both conformal and a massless minimal couplings. The latter requires certain regularization procedure in order to give the response function a physical meaning. We have discussed extension of these results to the de Sitter $\alpha$-vacua and have pointed out some possible ambiguities for a complex scalar. 

We have also shown that for a real scalar field theory with a field-detector coupling
linear in the field operator, with the interpretation discussed at the end of \ref{setup}, we can  compute the response function for the $\alpha$-vacua (see also \cite{Bousso:2001mw}). It is easy to argue that such computation extends to any arbitrary order of the perturbation theory. This is because at the $n$-th order, we have a term like  $ \prod_n \int d\tau_n \phi(\tau_n)$ from the $S$-matrix expansion. Since there are as many integrations as the number of field operators, we shall never have two Wightman functions appearing in a single integral. Accordingly, the cancellation of the $i \epsilon$
regulators as of \ref{alpha-com} does not occur in this case.  For example,  the second order correction in the response function can be evaluated (up to some numerical factors) as
$$\frac{d^2 {\cal F_{\alpha}}(\Delta E)}{ d t_+\,dt_+'} \sim  \left(\int_{-\infty }^{\infty} d(\Delta t)\, e^{- i \Delta E \Delta t } \,i G_{\alpha}^{+}(\Delta t)\right)^2\, +\,\int_{-\infty }^{\infty} d(\Delta t)\, e^{- i \Delta E \Delta t } \,i G_{\alpha}^{+}(\Delta t) \times \int_{-\infty }^{\infty} d(\Delta t')\, e^{ i \Delta E \Delta t' }\, i G_{\alpha}^{+}(\Delta t'),$$
which can indeed be computed without any ambiguity. Similarly, the corrections to the response function can be obtained  for higher orders as well.

For the massless and minimal complex scalar in particular (\ref{qcs2}), we have discussed a  viable regularisation cum renormalisation scheme. It will  be interesting to check how this can consistently tackle the divergences  at higher order of the perturbation theory  as well. 

Computation of the response function for a  fermionic field will be more realistic. Since massless fermion is conformally invariant, we may expect in this case the spectrum to be qualitatively similar to that of a complex conformal scalar. Also, we do not expect any de Sitter breaking logarithms for fermions.  It will also be interesting to investigate the massless minimal complex scalar field theory from various perspective  of quantum entanglement, e.g. entanglement harvesting. The effect of background primordial electromagnetic fields on a charged scalar will also be interesting, for in this case we expect  de Sitter breaking terms indicating instability at late times analogous to that of the growing logarithm of \ref{final}, even in the massive case. We shall come back to these issues in  future works.

\bigskip
\section*{Acknowledgements}
MSA's research is supported by the ISIRD grant 9-252/2016/IITRPR/708.  SB's research is partially supported by the ISIRD grant 9-289/2017/IITRPR/704. Research of KL is partially supported by the Department of Science and Technology (DST) of the Government of India through a research grant under INSPIRE Faculty Award (DST/INSPIRE/04/2016/000571). 
The authors would like to thank  S.~Chakrabortty, R.~Gupta and T.~Padmanabhan for useful discussions and critical comments. They also acknowledge anonymous referees for careful critical reading of the manuscript and for making valuable comments.

\bigskip

\end{document}